\documentclass[onecolumn]{emulateapj}
\usepackage{longtable}
\usepackage{graphicx}
\usepackage{graphics}
\usepackage{color}
\usepackage{epsfig}
\usepackage{rotating}
\usepackage{subfigure}
\usepackage{float}
\shortauthors{}
\shorttitle{}
\begin{document}

\title{\emph{Chandra} Survey of Nearby Highly Inclined Disc Galaxies - III:\\Comparison with Hydrodynamical Simulations of Circumgalactic Coronae}

\author{Jiang-Tao Li\altaffilmark{1,2,3}, Robert A. Crain\altaffilmark{4}, and Q. Daniel Wang\altaffilmark{2}} \altaffiltext{1}{Service d'Astrophysique (SAp)/IRFU/DSM/CEA Saclay, Bt. 709, 91191 Gif-sur-Yvette Cedex, France} \altaffiltext{2}{Department of Astronomy, University of Massachusetts, 710 North Pleasant Street, Amherst, MA 01003, U.S.A.} \altaffiltext{3}{Key Laboratory of Modern Astronomy and Astrophysics, Nanjing University, Ministry of Education, Nanjing, 210093, China} \altaffiltext{4}{Leiden Observatory, Leiden University, PO Box 9513, 2300 RA Leiden, Netherlands}


\nonumber

\begin{abstract}
X-ray observations of circumgalactic coronae provide a valuable means by which to test galaxy formation theories. Two primary mechanisms are thought to be responsible for the establishment of such coronae: accretion of intergalactic gas and/or galactic feedback. In this paper, we first compare our \emph{Chandra} sample of galactic coronae of 53 nearby highly-inclined disc galaxies to an analytical model considering only the accretion of intergalactic gas. We confirm the existing conclusion that this pure accretion model substantially over-predicts the coronal emission. We then select 30 field galaxies from our original sample, and correct their coronal luminosities to uniformly compare them to deep X-ray measurements of several massive disc galaxies from the literature, as well as to a comparable sample of simulated galaxies drawn from the Galaxies-Intergalactic Medium Interaction Calculation (GIMIC). These simulations explicitly model both accretion and supernovae feedback and yield galaxies that exhibit X-ray properties in broad agreement with our observational sample. However, notable and potentially instructive discrepancies exist between the slope and scatter of the $L_X-M_{200}$ and $L_X-{\rm SFR}$ relations, highlighting some known shortcomings of GIMIC, for example, the absence of AGN feedback, and possibly the adoption of constant stellar feedback parameters. The simulated galaxies exhibit a tight correlation (with little scatter) between coronal luminosity and halo mass. Having inferred $M_{200}$ for our observational sample via the Tully-Fisher relation, we find a weaker and more scattered correlation. In the simulated and observed samples alike, massive non-starburst galaxies above a typical transition mass of $M_*\sim2\times10^{11}\rm~M_\odot$ or $M_{200}\sim10^{13}\rm~M_\odot$ tend to have higher $L_X/M_*$ and $L_X/M_{200}$ than low-mass counterparts, indicating that the accretion of intergalactic gas plays an increasingly important role in establishing the observable hot circumgalactic medium with increasing galaxy mass.
\end{abstract}

\section{Introduction}\label{PaperIIIsec:Introduction}

The presence of extended X-ray emitting coronae associated with present-day massive disc galaxies is a generic prediction of cold dark matter (CDM) galaxy formation theory (e.g., \citealt{White91,Benson10}). The coronae serve as a reservoir of both diffuse, metal-poor gas accreted from the intergalactic medium (IGM), and metal-rich gas that is either ejected from galaxies by energetic feedback or stripped from infalling satellites. Two mechanisms are thought to be chiefly responsible for establishing the the X-ray emission properties of coronae: feedback injected by stellar wind/supernovae (SNe) (e.g., \citealt{Strickland00a}) and active galactic nuclei (AGN), and the shock heating and adiabatic compression of gas accreted from the IGM (e.g., \citealt{White78}).

Although galactic coronae have been detected around many disc galaxies, their origin remains uncertain. For example, the coronal luminosity is observed to correlate well with many tracers of feedback activity [e.g., the infrared (IR) or radio luminosities, \citealt{Strickland04b,Grimes05,Tullmann06b,Li08,Li13b}], and the heavy element abundance of the X-ray luminous gas is also indicative of enrichment by SNe (e.g., \citealt{Martin02,Ji09,Li09,Bregman13,Li13b}). These are often regarded as direct evidence for the feedback scenario (but see \citealt{Crain13} for an alternative interpretation). Furthermore, non-gravitational heating is also indicated by the change of slope of the scaling relations between the coronal luminosity/temperature and the galaxy mass in different mass ranges (e.g., \citealt{Ponman99,Ponman03,OSullivan03}). The uncertainty is driven by the difficulty of detecting and characterizing the diffuse thermal X-ray emission at large galactocentric radii (e.g., \citealt{Benson00,Rasmussen09}). In recent years, with either stacking of large numbers of X-ray observations of various types of galaxies \citep{Anderson13} or moderately deep single-pointing X-ray observations of massive disc galaxies \citep{Anderson11,Dai12,Bogdan13a,Bogdan13b}, X-ray emission from the putative coronae on large scales (at radii up to $\sim50\rm~kpc$) has been reported around several galaxies which are not particularly active in SF. These coronae are considered to be most likely produced by the accretion of intergalactic gas. However, such detections are still rare and only concern galaxies within a narrow range of stellar mass (see Table~\ref{table:GalaxyPara}). Comparison with X-ray measurements of galaxies in a broad mass range is thus particularly desirable.

In our previous studies \citep{Li13a,Li13b} (hereafter, Paper~I, II), we investigated the galactic coronae associated with a sample of \emph{Chandra}-observed nearby highly-inclined disc galaxies. We found a tight correlation between the coronal luminosity and the SN energy input rate; the abundance ratio of the coronal gas is also consistent with a combined contribution from core collapsed and Type~Ia SNe. These results can be qualitatively interpreted, naively, as evidence that galactic coronae are mainly established by SN feedback. However, our sample only includes low and intermediate mass galaxies (typically with stellar mass $M_*\lesssim 2\times10^{11}\rm~M_\odot$), which may not be massive enough for accreted gas to contribute a dominant fraction of the detected X-ray flux. In fact, recent cosmological hydrodynamical simulations invoking both accretion and feedback predict a coronal luminosity range that is broadly consistent with X-ray observations \citep{Crain10a}. These simulations trace the dynamics of the coronal gas, a large fraction of which is quasi-hydrostatic or even infalling. \citet{Crain13} further showed that the high ($\sim$solar) metallicities inferred from the X-ray spectroscopy are not in conflict with models in which the bulk (by mass) of the hot circumgalactic medium (CGM) is established by accretion. The soft X-ray emission is dominated by collisionally-excited metal ions deposited by feedback, whilst the majority of the accreted gas is metal poor and is comparatively radiatively inefficient. The hot CGM thus appears to be typically metal-poor in a mass-weighted sense, but its X-ray luminosity-weighted metallicity is often close to solar. Therefore, accretion of intergalactic gas may still be important, especially in those massive galaxies not included in our \emph{Chandra} sample. In Paper~II, we also compared our measurements to X-ray observations of massive elliptical galaxies, which have significantly steeper $L_X-M_*$ relations. The formation of such galaxies may, however, be subject to more violent mergers and the corresponding starbursting process, potentially complicating the interpretation of observations of their coronae. Furthermore, the typically rich clustered environment of massive elliptical galaxies also potentially contaminates the measurement of coronal properties associated with individual galaxies. Therefore, comparisons with X-ray measurements of elliptical galaxies do not offer the cleanest means by which to place direct constraint on disc galaxy formation models.

We herein conduct a quantitative comparison of our X-ray measurements of disc galaxies to the X-ray measurements of several isolated massive disc galaxies from the literature (\S\ref{PaperIIIsubsec:MassiveGalaxy}), as well as the results of hydrodynamical simulations invoking both accretion and feedback (i.e., the GIMIC simulations, see \S\ref{PaperIIIsubsec:Simulations}), in order to further explore the origin of coronae around disc galaxies and place direct constraint on galaxy formation models. This paper is organized as follows. In \S\ref{PaperIIIsec:data}, we describe our procedure for correcting coronal luminosity measurements to ensure data homogenization between the galaxies selected from our original \emph{Chandra} sample, galaxies from the literature, and the GIMIC simulations. We compare the observational results to an analytical model as well as the GIMIC simulations in \S\ref{PaperIIIsec:Comparison}, and further discuss the scientific implications in \S\ref{PaperIIIsec:Discussion}. Our results and conclusions are summarized in \S\ref{PaperIIIsec:Summary}.

\section{Data Homogenization of Observations and Simulations}\label{PaperIIIsec:data}

In order to compare measurements from different observations and simulations in a uniform fashion, corrections to ensure the X-ray measurements for data homogenization are necessary. We aim to compare the coronal luminosity measured in a consistent manner, i.e., in the same band and in the same radial range.

\subsection{X-ray Measurements from the Chandra Sample}\label{PaperIIIsubsec:OurSample}

Here, we consider only field galaxies from our original \emph{Chandra} sample (Paper~I), because X-ray measurements of the clustered galaxies are potentially contaminated by the intracluster medium, and also because the massive disc galaxies whose measurements we collate from the literature (\S\ref{PaperIIIsubsec:MassiveGalaxy}) and the selected galaxies from the GIMIC simulations (\S\ref{PaperIIIsubsec:Simulations}) are in the field. We also select galaxies with published rotation velocity measurements (inclination-corrected, $v_{rot}$, as listed in Table~\ref{table:GalaxyPara}), which are used for estimating the halo masses and radii (see below). One galaxy, NGC~3384, has extremely low $v_{rot}$ ($\sim17\rm~km~s^{-1}$); this is likely a result of interaction with its companions, and hence an inaccurate reflection of the depth of its gravitational potential (Paper~I). This galaxy is thus excluded from the current sample. In total, 30 galaxies are selected and their key parameters are summarized in Table~\ref{table:GalaxyPara}.

%\clearpage
\begin{deluxetable}{lccccccccccc}
\centering
\tiny %\tiny\scriptsize\footnotesize\small\normalsize\large\Large\LARGE\huge\Huge
%\ptlandscape
  \tabletypesize{\tiny}
  \tablecaption{Parameters of Galaxies}
  \tablewidth{0pt}
  \tablehead{
\colhead{Name} & \colhead{$\log M_*$} & \colhead{SFR} & \colhead{$v_{rot}$} & \colhead{$\log M_{200}$} & \colhead{$r_{200}$} & \colhead{$\log L_X$} \\
   & ($\rm M_\odot$) & ($\rm M_\odot yr^{-1}$) & ($\rm km~s^{-1}$) & ($\rm M_\odot$) & (kpc) & ($\rm erg/s$)
}
\startdata
IC2560 & 10.03 & $2.05\pm0.32$ & $196\pm3$ & $12.10\pm0.02$ & 225 & $39.70\pm0.03$ \\
M82 & 10.30 & $7.70\pm0.46$ & $100\pm10$ & $11.16\pm0.14$ & 108 & $39.56\pm0.002$ \\
NGC0024 & 9.18 & $0.11_{-0.06}^{+0.01}$ & $93\pm1$ & $11.07\pm0.02$ & 101 & $37.81\pm0.08$ \\
NGC0520 & 10.57 & $11.74\pm1.60$ & $72\pm2$ & $10.70\pm0.05$ & 76 & $39.11_{-0.14}^{+0.08}$ \\
NGC0660 & 10.47 & $7.13\pm0.90$ & $140\pm3$ & $11.64\pm0.03$ & 157 & $38.35_{-0.08}^{+0.11}$ \\
NGC0891 & 10.69 & $2.46\pm0.34$ & $212\pm5$ & $12.22\pm0.03$ & 245 & $38.83\pm0.01$ \\
NGC1023 & 10.83 & - & $112\pm5$ & $11.33\pm0.07$ & 124 & $37.84_{-0.11}^{+0.09}$ \\
NGC1482 & 10.36 & $6.53\pm0.78$ & $121\pm8$ & $11.43\pm0.10$ & 133 & $39.33\pm0.04$ \\
NGC1808 & 10.58 & $7.78\pm0.62$ & $122\pm5$ & $11.44\pm0.06$ & 135 & $38.50_{-0.05}^{+0.04}$ \\
NGC2787 & 10.52 & $0.26_{-0.23}^{+0.003}$ & $181\pm13$ & $12.00\pm0.10$ & 207 & $36.86_{-0.33}^{+0.45}$ \\
NGC2841 & 10.99 & $0.49_{-0.14}^{+0.05}$ & $318\pm9$ & $12.79\pm0.04$ & 379 & $38.01_{-0.07}^{+0.06}$ \\
NGC3079 & 10.47 & $6.06\pm0.58$ & $210\pm5$ & $12.20\pm0.03$ & 242 & $39.63\pm0.02$ \\
NGC3115 & 10.83 & - & $107\pm5$ & $11.26\pm0.07$ & 117 & $36.21_{-0.20}^{+0.13}$ \\
NGC3198 & 10.02 & $0.62_{-0.12}^{+0.06}$ & $148\pm4$ & $11.71\pm0.04$ & 166 & $38.80\pm0.05$ \\
NGC3521 & 10.85 & $2.13_{-0.40}^{+0.24}$ & $233\pm6$ & $12.35\pm0.04$ & 271 & $38.87\pm0.03$ \\
NGC3556 & 10.21 & $1.57\pm0.15$ & $153\pm2$ & $11.76\pm0.03$ & 172 & $38.36_{-0.09}^{+0.08}$ \\
NGC3628 & 10.83 & $4.83\pm0.54$ & $215\pm3$ & $12.23\pm0.02$ & 248 & $39.47_{-0.04}^{+0.03}$ \\
NGC3955 & 10.11 & $2.08\pm0.27$ & 86 & 10.95 & 93 & $38.65_{-0.25}^{+0.15}$ \\
NGC4244 & 8.95 & $0.02_{-0.01}^{+0.001}$ & $89\pm2$ & $11.00\pm0.03$ & 96 & $37.94\pm0.05$ \\
NGC4594 & 11.19 & $0.28_{-0.10}^{+0.03}$ & $358\pm10$ & $12.95\pm0.04$ & 430 & $38.92_{-0.03}^{+0.02}$ \\
NGC4631 & 10.01 & $1.67\pm0.20$ & $138\pm3$ & $11.62\pm0.04$ & 155 & $39.10\pm0.02$ \\
NGC4666 & 10.61 & $4.04\pm0.45$ & $193\pm2$ & $12.08\pm0.02$ & 221 & $39.08_{-0.19}^{+0.07}$ \\
NGC5102 & 9.18 & $0.01_{-0.005}^{+0.0005}$ & 90 & 11.01 & 97 & $36.75\pm0.09$ \\
NGC5170 & 10.67 & $0.56_{-0.29}^{+0.04}$ & 245 & 12.41 & 285 & $38.88\pm0.11$ \\
NGC5253 & 8.69 & $0.34\pm0.03$ & $38\pm1$ & $9.81\pm0.06$ & 38 & $37.64\pm0.03$ \\
NGC6503 & 9.50 & $0.15\pm0.01$ & $66\pm1$ & $10.60\pm0.04$ & 70 & $37.48_{-0.06}^{+0.05}$ \\
NGC6764 & 10.00 & $2.73\pm0.25$ & $139\pm3$ & $11.63\pm0.04$ & 156 & $38.05_{-0.18}^{+0.13}$ \\
NGC7090 & 9.17 & $0.14_{-0.03}^{+0.01}$ & 102 & 11.19 & 111 & $36.40_{-0.38}^{+0.32}$ \\
NGC7582 & 10.83 & $12.84\pm1.69$ & $194\pm3$ & $12.10\pm0.02$ & 223 & $38.49_{-0.08}^{+0.07}$ \\
NGC7814 & 10.85 & - & $230\pm8$ & $12.33\pm0.05$ & 268 & $37.95\pm0.07$ \\
\hline
NGC266 & 11.30 & 2.4 & $868\pm50$ & 12.90 & 410 & $40.47\pm0.40$ \\
NGC1961 & 11.62 & 15.5 & $447\pm14$ & 13.08 & 470 & $40.70\pm1.81$ \\
NGC6753 & 11.51 & 11.8 & $395\pm21$ & 13.00 & 440 & $40.88\pm1.42$ \\
UGC12591 & 11.91 & 4.8 & $488\pm12$ & 13.38 & 601 & $40.53\pm0.36$
\enddata
\tablecomments{\scriptsize Parameters of galaxies from our \emph{Chandra} sample (Paper~I; those above the solid line) and the literature (those below the solid line): $M_*$ is the stellar mass estimated using the K-band luminosity from the 2MASS extended source catalogue \citep{Skrutskie06}; SFR is the star formation rate estimated in the same way as Paper~I using the \emph{IRAS} data; $v_{rot}$ is the maximum rotation velocity corrected for inclination, obtained from the HyperLeda database (http://leda.univ-lyon1.fr/); $M_{200}$ is the dark matter halo mass defined as the mass within a sphere with a mean density of $200\rho_{crit}$ (with a radius of $r_{200}$), where $\rho_{crit}$ is the critical density of the Universe. The extinction-corrected coronal luminosity ($L_X$) is measured in 0.5-2~keV and within $0.01-0.1~r_{200}$, as detailed in the text.
}\label{table:GalaxyPara}
\end{deluxetable}

We first estimate the virial mass and radius of each dark matter halo. We represent these quantities by $M_{200}$ and $r_{200}$, respectively, which are the total mass and radius of the halo with a mean density of $200\rho_{crit}$ [we use the Hubble constant from the \emph{WMAP} 9-years data \citep{Bennett13} to calculate $\rho_{crit}$, which is the critical density of the Universe]. We estimate $M_{200}$ using the scaling relation applicable to dark matter halos from \citet{Navarro97} ($M_{200}\propto v_{rot}^{3.23}$). $r_{200}$ is computed using the definition of $M_{200}$: $M_{200}=200\rho_{crit}\frac{4\pi}{3}r_{200}^3$. $M_{200}$ and $r_{200}$ of the sample galaxies are listed in Table~\ref{table:GalaxyPara}.

For consistency, we measure the coronal properties within a radial range that scales with the galaxy mass or $r_{200}$. The lower limit of the range should be large enough to exclude the X-ray emission from the galactic disc, while the upper limit should match or not significantly exceed the scale accessed by the \emph{Chandra} observations. For the typical coronal size of our \emph{Chandra} sample (Paper~I), we choose this range to be $0.01-0.1r_{200}$. This range also covers a significant fraction of the X-ray emission of coronae in hydrodynamical simulations  (e.g., \citealt{Crain10a,Crain13}).

We renormalize the coronal luminosity measured within a typical vertical range of $\pm 5h_{exp}$ ($h_{exp}$ is the exponential scale height of the 0.5-1.5~keV diffuse X-ray intensity profile) and radial range of $\pm D_{25}/4$ ($D_{25}$ is the B-band diameter of the projected major axis at the isophotal level of $25\rm~mag~arcsec^{-2}$) by extrapolating the intensity profile with an exponential model characterized by the diffuse X-ray vertical and radial scale heights of the sample galaxies (Paper~I). The exponential model used to fit the intensity profiles potentially inaccurately describes the X-ray intensity distribution at large radii, but the choice of $0.1r_{200}$ as the upper limit of the radii range (typically not exceeding $\pm 5h_{exp}$ and $\pm D_{25}/4$) minimizes the uncertainty resulting from inaccurate exponential fits. The renormalized coronal luminosities of the sample galaxies are listed in Table~\ref{table:GalaxyPara}, with the original statistical errors from the coronal luminosity measurements (Paper~I).

\subsection{Massive Disc Galaxies from the Literature}\label{PaperIIIsubsec:MassiveGalaxy}

Our \emph{Chandra} survey does not include galaxies with $M_*\gtrsim 2\times10^{11}\rm~M_\odot$. These massive disc galaxies are rare in nature, and consequently no such galaxy satisfies the selection criteria established in Paper~I. Several recent studies, however, report the detection of large-scale diffuse X-ray emission associated with massive disc galaxies in the local Universe (e.g., \citealt{Anderson11} for \emph{Chandra} observations of NGC~1961, \citealt{Dai12} for \emph{XMM-Newton} observations of UGC~12591, \citealt{Bogdan13a} for \emph{XMM-Newton} observations of NGC~1961 and NGC~6753, and \citealt{Bogdan13b} for \emph{ROSAT} and \emph{Chandra} observations of NGC~266). We supplement our sample with these measurements, applying appropriate corrections to their coronal luminosities to ensure a uniform comparison.

$M_{200}$ and $r_{200}$ of NGC~1961, NGC~6753, and UGC~12591 are directly taken from the above cited references or estimated from the inclination-corrected rotation velocities, in the same way as for our \emph{Chandra} sample galaxies (\S\ref{PaperIIIsubsec:OurSample}). However, the inclination angle of NGC~266 ($\sim14.5^\circ$) is too low to enable a reliable estimate of the rotation velocity (the apparent maximum rotation velocity of gas is $\sim218\rm~km~s^{-1}$, c.f the inclination corrected $\sim868\rm~km~s^{-1}$). We therefore use the halo mass estimated from the stellar mass of the galaxy, as adopted in \citet{Bogdan13b}.

For NGC~266, NGC~1961 (for which the results from \citealt{Bogdan13a} are used throughout the present paper), and NGC~6753, the coronal luminosities are measured in the 0.5-2~keV band, same as for our \emph{Chandra} sample. For UGC~12591, however, the original coronal luminosity from \citet{Dai12} is measured in the 0.6-1.4~keV band. We then adopt the spectral model described in \citet{Dai12} ($kT=0.64\rm~keV$, 0.5 solar abundance, and Galactic foreground absorption column density) to estimate the luminosity in the 0.5-2~keV band. This correction factor is only of order $\sim10\%$. We also adopt a redshift-independent distance of $134\rm~Mpc$ to UGC~12591 (from the NED) instead of the artificial 100~Mpc distance taken in \citet{Dai12}. Relevant galaxy parameters ($M_*$, SFR, and $L_X$) are also corrected for this revised distance estimate.

The largest correction factor is for the radial range adopted in measuring the coronal luminosity. According to the cited references, this range is $0.05-0.15r_{200}$ for NGC~266 \citep{Bogdan13b}, NGC~1961, and NGC~6753 \citep{Bogdan13a}, but is within 50~kpc for UGC~12591 \citep{Dai12}. We use the model ($\beta$-model or modified $\beta$-model) adopted in fitting the radial intensity profiles as described in the references to renormalize the coronal luminosity to our adopted $0.01-0.1r_{200}$ range. However, the published \emph{ROSAT} and \emph{Chandra} observations of NGC~266 are insufficiently deep for such a radial profile analysis. We therefore use the ratio of the count rate within the $0.01-0.1r_{200}$ and $0.05-0.15r_{200}$ ranges (after subtracting the contribution from the scattered photons of the central low luminosity AGN) to adopt a rough correction (private communication with $\rm\acute{A}kos$ Bogd$\rm\acute{a}$n). We also subtract the estimated contributions from stellar sources [bright low mass X-ray binaries (LMXBs), \citealt{Gilfanov04}; faint LMXBs, cataclysmic variables (CVs) and coronal active binaries (ABs), \citealt{Revnivtsev08}; high mass X-ray binaries (HMXBs), \citealt{Mineo12}] within $0.01-0.1r_{200}$ of NGC~266, NGC~1961, and NGC~6753, by scaling from their stellar mass and SFR. These stellar source contributions are not accounted for in the original references. The final corrected coronal luminosities are listed in Table~\ref{table:GalaxyPara}.

\subsection{Numerical Simulations}\label{PaperIIIsubsec:Simulations}

We use the results from the Galaxies-Intergalactic Medium Interaction Calculation (GIMIC; \citealt{Crain09}), a suite of hydrodynamic resimulations of regions drawn from the Millennium simulation \citep{Springel05b}. The GIMIC simulations are performed with a variant of the TreePM-SPH code GADGET3, a substantial upgrade of GADGET2 \citep{Springel05a}. GIMIC follows the evolution of five representative roughly spherical regions with different overdensities drawn from the dark matter Millennium simulation \citep{Springel05b}. The GIMIC simulations have intermediate ($m_{gas}=1.16\times10^7 h^{-1}\rm M_\odot$) and high resolution ($m_{gas}=1.45\times10^6 h^{-1}\rm M_\odot$), compared to the ``low'' resolution for the original Millennium simulation, in which the collisionless particles, representing a composite of baryonic and dark matter, have mass of $8.6\times10^8 h^{-1}\rm M_\odot$. In the present paper, we use the intermediate resolution simulations, all of which are run to z=0. \citet{Crain10a,Crain13} demonstrate that the properties explored here are numerically converged in the regimes of interest, and that the resolution is sufficient to enable morphological classification of $\sim L^\star$ galaxies.

Metals typically dominate the specific emissivity of astrophysical plasmas (e.g., \citealt{Smith01,Wiersma09}). Therefore, the accurate prediction of coronal X-ray luminosities requires that simulations, besides modelling the gravitational and thermodynamical evolution of gas, must track the nucleosynthesis of metal species by stellar populations, and macroscopic transport of this material throughout the cosmological growth and assembly of galaxies (and groups and clusters of galaxies). These processes are modeled in GIMIC, and have been shown to play a key role in the establishment of the amplitude, and radial profile of, the luminosity and metallicity of coronae \citep{Crain13}. One shortcoming of GIMIC, however, is that it does not model the evolution of black holes or feedback effects associated with them. Therefore, we caution that the stellar mass and the radiative cooling (and hence the coronal X-ray luminosity) of the most massive galaxies in the simulated sample (where AGN feedback is expected to be non-negligible) may be inadequately modeled (see \S\ref{PaperIIIsubsec:AGNfeedback} for further discussion).

We adopt the same disc galaxy sample as \citet{Crain10a}, selecting central galaxies within friends-of-friends (FoF) halos, and focussing on $\sim L^\star$ disc galaxies by requiring a disc-to-total stellar mass ratio $D/T>0.3$ and a stellar mass in the range $10^{10}<M_*<10^{11.7}\rm M_\odot$. \citet{Crain10a} considered only isolated systems by excluding galaxies that are interacting or are members of galaxy groups and clusters. As per \citet{Crain10a}, the X-ray luminosity of a corona is computed by summing the luminosities of all gas particles bound to the corresponding subhalo. The cooling function is computed using the Astrophysical Plasma Emission Code (APEC, v1.3.1, \citealt{Smith01}) under the assumption that the gas is in collisional ionization equilibrium. As is generally the case for cosmological simulations, the adopted resolution is insufficient to model the formation of individual stars within a multi-phase interstellar medium. The simulations therefore impose a polytropic equation of state to the high density ($n_H>0.1\rm~cm^{-3}$) gas that is subject to thermo-gravitational instability (e.g. \citealt{Schaye04,Schaye08}). This gas (mostly in the disc) is assigned a temperature of that is below the minimum considered by the APEC cooling tables; therefore, by construction, the computed luminosity comprises only extraplanar emission. We herein calculate the coronal luminosity within $0.01-0.1r_{200}$, which is dominated by hot gas and less affected by the multi-phase gas in the disc. To renormalize the original coronal X-ray luminosities from \citet{Crain10a}, which were measured for the entire halo to $r=0.01-0.1r_{200}$, we compute the average stacked radial intensity profiles in five halo mass bins [$\log (M_{200}/M_\odot)=11.75-12.00, 12.00-12.25, 12.25-12.50, 12.50-12.75, 12.75-13.00$], and then calculate the luminosity fraction enclosed within $0.01-0.1r_{200}$. This fraction is typically $\sim(20-50)\%$ of the total coronal luminosity. We corrected this enclosed fraction to compute the 0.5-2~keV luminosity within $0.01-0.1r_{200}$ from the original total coronal luminosity. The SFR (stellar mass) of a simulated galaxy is estimated by summing the SFRs (stellar masses) of all gas (star) particles bound to a galaxy's subhalo. The rotation velocity, $v_{rot}$, is obtained from the maximum value of the halo velocity curve, assuming spherical symmetry.

\section{Comparison between Observation and Theory}\label{PaperIIIsec:Comparison}

\subsection{Analytical model involving only gravitational heating}\label{PaperIIIsubsec:analytical}

Before comparing our X-ray measurements to results from GIMIC, it is instructive to consider a simple analytical model involving only gravitational energy release of the accreted gas (no feedback) as a reference for the comparison. The inaccuracy of this simple accretion-only model \citep{Benson00}, as compared to GIMIC, was highlighted in \citet{Crain10a}. In this subsection, we adopt the uniformly corrected coronal luminosity from Papers~I and II, without any further corrections as described in \S\ref{PaperIIIsubsec:OurSample}, in order to include the X-ray emission from the galactic disc. Quantitative comparisons with X-ray measurements of massive disc galaxies and the results from GIMIC will be presented in the following sections.

\begin{figure}[!h]
\begin{center}
\epsfig{figure=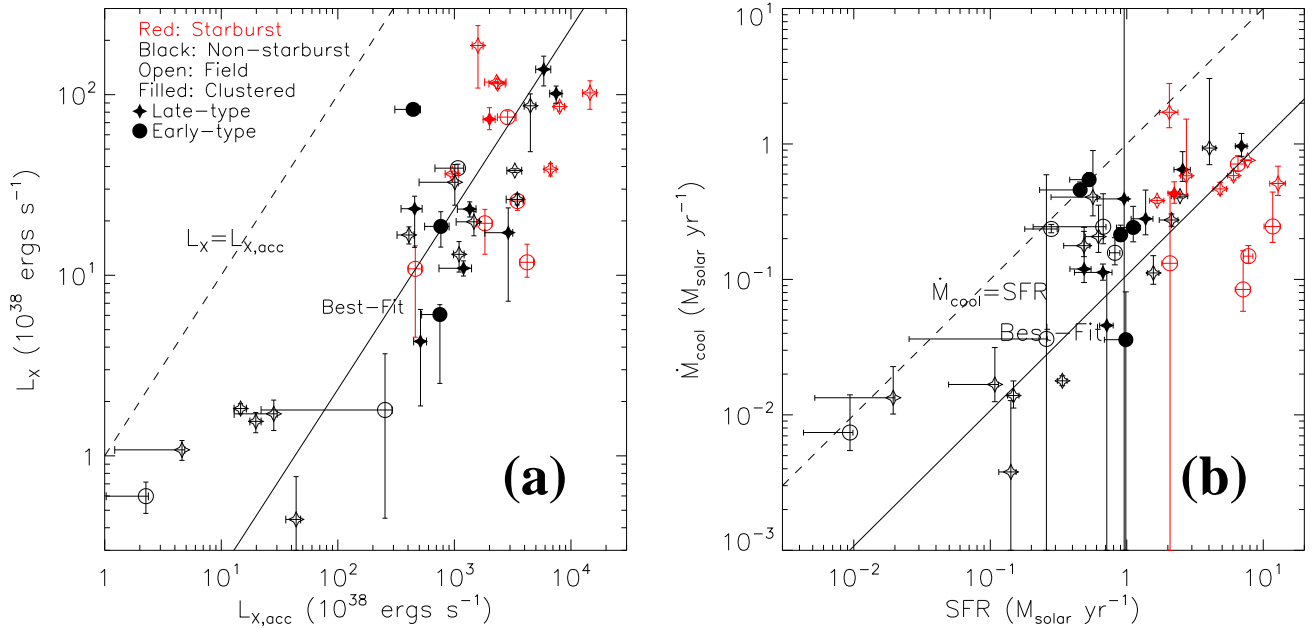,width=1.0\textwidth,angle=0, clip=}
\caption{Comparison of the X-ray measurements to the predictions from an analytical model as described in \S\ref{PaperIIIsubsec:analytical}. (a) The measured coronal luminosities of our \emph{Chandra} sample ($L_X$) plotted against the coronal luminosity expected from the accretion only model ($L_{X,acc}$). The solid line shows the best-fit linear relation (Eq.~\ref{equi:AccretionModelObservation}), while the dashed line marks $L_X=L_{X,acc}$. All the X-ray measurements are obtained from our \emph{Chandra} observations (Papers~I and II) without further corrections as described in \S\ref{PaperIIIsubsec:OurSample}; different colors and symbols denote various star formation, environmental, and morphological subclasses of the galaxies as classified in Paper~I. (b) The radiative cooling rate of the X-ray emitting coronal gas ($\dot{M}_{cool}$) vs. the SFR of the galaxies. The solid line shows the best-fit linear relation (Eq.~\ref{equi:RcoolSFR}), while the dashed line shows where $\dot{M}_{cool}=\rm SFR$. Symbols are the same as those in panel~(a).}\label{fig:analyticalmodel}
\end{center}
\end{figure}

We estimate the expected coronal X-ray luminosity ($L_{X,acc}$), in the absence of any effects of feedback on the mass, structure and emissivity of the hot CGM, following the procedure detailed in \citet{Benson00}. Assuming the accretion rate $\dot{M}_{acc}$ can be connected to the SFR by the poorly constrained SF efficiency ($\zeta={\rm~SFR}/\dot{M}_{acc}$; e.g., \citealt{Martin06,Dave11}), $L_{X,acc}$ can be predicted with the observational parameters $v_{rot}$ and SFR as:
\begin{equation}\label{equi:ModelLX}
L_{X,acc}\sim4.5 {\rm SFR} v_{rot}^2/\zeta.
\end{equation}
Assuming $\zeta\sim1$, we then compare $L_{X,acc}$ to the observed X-ray luminosity from our measurements (Fig.~\ref{fig:analyticalmodel}a) and obtain a best-fit linear relation of (see \S3.1 of Paper~II for the method of fitting):
\begin{equation}\label{equi:AccretionModelObservation}
L_X=10^{-(1.63\pm0.08)}L_{X,acc}.
\end{equation}
The relation is just qualitative, showing the general correlation and large inconsistency between the gravitational energy and the X-ray luminosity. The coefficient is, clearly, far below unity ($\sim2\%$), consistent with conclusions reached previously (e.g., \citealt{White91,Benson00,Toft02}). Reducing $\zeta$ to a value significantly below unity (as suggested by many works; e.g., \citealt{Dave11}) further increases this discrepancy, while small changes of the coefficient in Eq.~\ref{equi:ModelLX} are insufficient to compensate it (e.g., \citealt{White91} adopted a value of 2.5). The small photometry region of our sample galaxies also seems insufficient to explain this large discrepancy, because the inner region of a halo often dominates the X-ray emission owing to the high density and metallicity of the hot gas \citep{Crain13}.

We further highlight the oversimplicity of the pure accretion model by comparing the halo gas radiative cooling rate ($\dot{M}_{cool}$; Paper~I) to the SFR inferred from the IR luminosity. As shown in Fig.~\ref{fig:analyticalmodel}b, the $\dot{M}_{cool}-{\rm SFR}$ relation can be characterized with a linear function:
\begin{equation}\label{equi:RcoolSFR}
\dot{M}_{cool}=11.4(<29.5)\%{\rm SFR},
\end{equation}
indicating that the radiative cooling of the X-ray emitting corona accounts for only a small fraction of the current SFR. This is a remarkably different situation from observed in many massive elliptical galaxies, for which the radiative cooling rate is often far greater than the current SFR (e.g., \citealt{Mathews03}).

The above analytical model assumes that only the accretion of gas that has cooled via X-ray emission in the vicinity of galactic discs replenishes the gas for the disc SF. In reality, other fueling mechanisms, such as the accretion without cooling via soft X-ray emission, or the recycling of gas from existing stellar populations, are likely efficient to replenish the cool gas consumed in SF. In particular, \citet{Leitner11} found that the recycled gas from stellar mass loss can provide most or all of the fuel required to sustain the current level of SF in late-type galaxies. In addition, gas accretion is thought to be bimodal, with maximum past temperatures either of order the virial temperature or $\lesssim10^5\rm~K$ (e.g., \citealt{Keres05,vandeVoort11}). In general, the cold mode accretion (at a low temperature) is believed to dominate the growth of low mass galaxies with baryonic mass typically $\lesssim10^{10.3}\rm~M_\odot$ \citep{Keres05}. Furthermore, \citet{vandeVoort11} have found that for the accretion onto galaxies (as opposed to accretion onto halos), cold mode is always significant and the majority of stars present in any mass halos at any redshifts were formed from gas accreted in cold mode.

The discrepancy between the X-ray radiative cooling rate and the current SFR (Eq.~\ref{equi:RcoolSFR}) apparently indicates a higher than unity SF efficiency $\zeta$, which further indicates the existence of additional gas sources as discussed above. But even if we consider these gas sources (i.e., by substituting Eq.~\ref{equi:RcoolSFR} into Eq.~\ref{equi:ModelLX}), Eq.~\ref{equi:ModelLX} still tends to over-predict the coronal luminosity by a factor of five. It is thus obvious that the pure accretion model \emph{cannot} reproduce the observed X-ray luminosity, as previously pointed out by \citet{Crain10a} and others (e.g., \citealt{Benson00}).

\subsection{Comparison between Observations and Simulations}\label{PaperIIIsubsec:compareGIMIC}

We next compare the X-ray measurements of our \emph{Chandra} sample and the massive disc galaxies from the literature (\S\ref{PaperIIIsubsec:MassiveGalaxy}) to the results from the GIMIC simulations, which explicitly model both the accretion of gas in a cosmological framework and the feedback that results from the formation and evolution of stars. Here we are careful to compare the coronal luminosity ($L_X$) of the observed and simulated galaxies within the same X-ray bandpass and the same radial aperture size (\S\ref{PaperIIIsec:data}). A quantitative and uniform characterization of the thermal and chemical states (temperature and metal abundance) of the coronal gas is significantly more complicated. Readers interested in such issues may refer to \citet{Crain13} or our discussions in Papers~I and II.

\begin{figure}[!h]
\begin{center}
\epsfig{figure=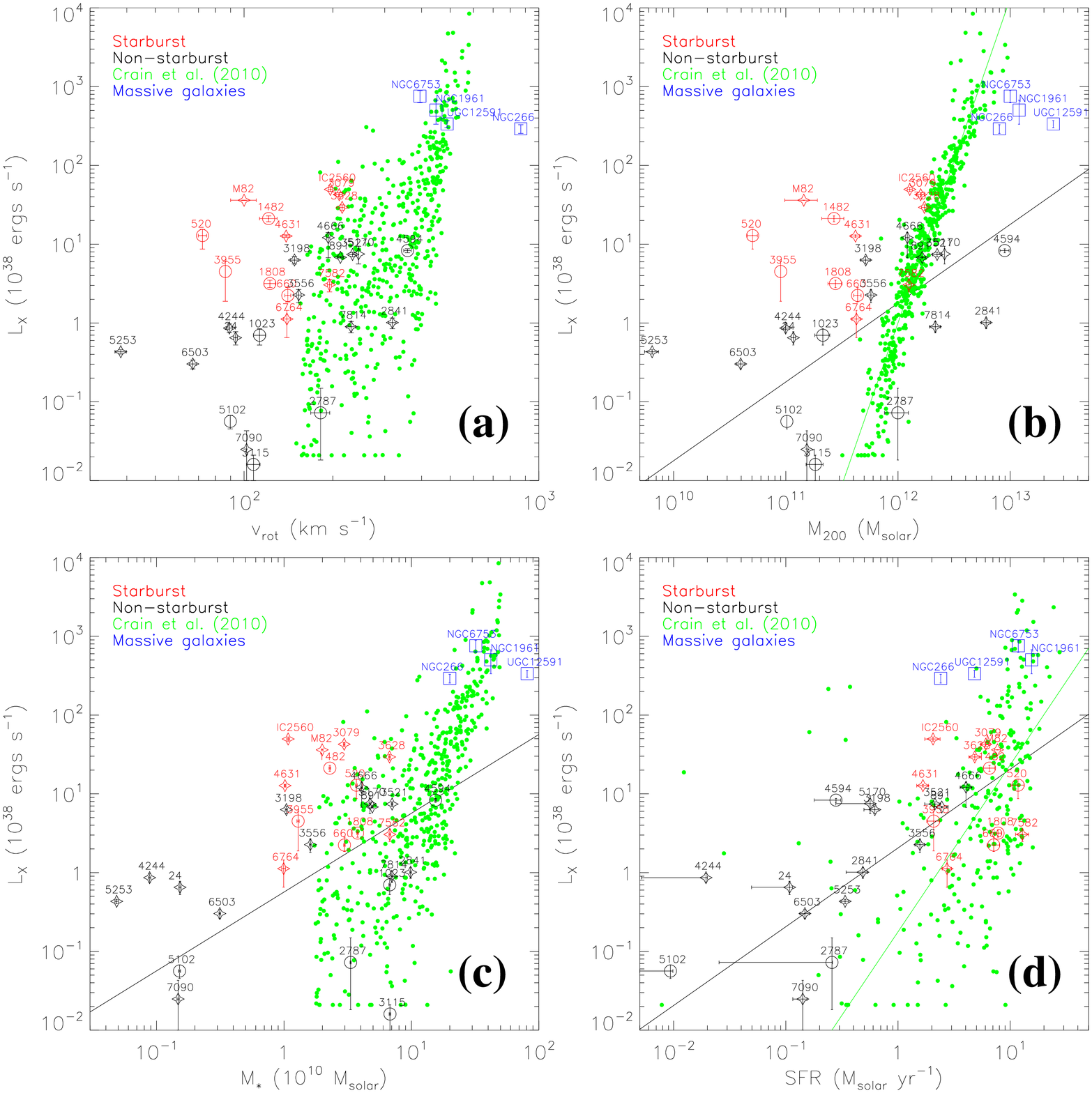,width=1.0\textwidth,angle=0, clip=}
\caption{Comparisons of the X-ray observations to the results from the GIMIC simulations. (a) $L_X$ v.s. the inclination-corrected maximum gas rotation velocity $v_{rot}$. (b) $L_X$ v.s. $M_{200}$. (c) $L_X$ v.s. $M_*$. (d) $L_X$ v.s. SFR. Black and red symbols are the same as those in Fig.~\ref{fig:analyticalmodel}; blue boxes denote the massive disc galaxies listed in Table~\ref{table:GalaxyPara}; green dots denote the simulated galaxies from GIMIC. The solid lines in panels~(b) and (c) are the linear relations fitted to the non-starburst field galaxies of our \emph{Chandra} sample (the black symbols), while that in panel~(d) is fitted to both the starburst and non-starburst galaxies of our \emph{Chandra} sample (the black and red symbols). The green lines in panels (b) and (d) are the power laws fitted to the GIMIC galaxies (in panel d, we only use the galaxies with ${\rm SFR}>0.5\rm~M_\odot~yr^{-1}$).}\label{fig:compareGIMIC}
\end{center}
\end{figure}

As shown in Fig.~\ref{fig:compareGIMIC}, $L_X$ of the GIMIC galaxies better matches the X-ray observations than the simple accretion model predictions of \S\ref{PaperIIIsubsec:analytical}. In fact, GIMIC reproduces not only the coronal X-ray emission in $0.01-0.1r_{200}$ of $L^\star$ galaxies, but the scatter in $L_X$ at a given $v_{rot}$ or $M_*$ as well (Fig.~\ref{fig:compareGIMIC}a,c). The substantial reduction of the coronal luminosity in the GIMIC simulations as compared to the pure accretion model is due to the mass reduction of hot gas within a dark matter halo, as a result of the selective cooling and dropout of low entropy gas (e.g., via SF) and galactic feedback \citep{Crain10a}.

The observed and simulated galaxies generally overlap with each other in the $L_X-v_{rot}$ plane (Fig.~\ref{fig:compareGIMIC}a). The most significant outlier is NGC~266, for which the inclination correction of the rotation velocity is poorly constrained (\S\ref{PaperIIIsubsec:MassiveGalaxy}). Furthermore, starburst galaxies in our \emph{Chandra} sample (the red data points) also appear to be systematically more X-ray luminous (at a given $v_{rot}$) than non-starburst ones, as well as the GIMIC galaxies.

The dark matter halo dominates the potential on large scales in a galaxy. \citet{Kim13} have recently found a tighter correlation of the coronal luminosity with the total mass (than with the stellar mass) for a sample of early-type galaxies. This result is qualitatively consistent with the $L_X-v_{rot}$ correlation shown in Fig.~\ref{fig:compareGIMIC}a, although for different types of galaxies. This consistency indicates, as is clearly implied by GIMIC, that the halo mass may be a primary factor in the retention of hot gas in general. Therefore, we further convert the rotation velocity to the dark matter halo mass for each of our observed sample galaxies, using the $M_{200}-v_{rot}$ relation of \citet{Navarro97} (\S\ref{PaperIIIsec:data}) and plot $L_X$ v.s. $M_{200}$ in Fig.~\ref{fig:compareGIMIC}b. This halo mass estimation differs from those using the stellar mass-halo mass relation (SMHMR) obtained from the abundance matching technique (e.g., \citealt{Behroozi10}), especially at the high mass end where the large halo mass in SMHMR is mostly for groups/clusters of galaxies (Fig.~\ref{fig:M200Scaling}a).

Since the starburst galaxies in our \emph{Chandra} sample and the four massive disc galaxies from the literature appear to be significantly more X-ray luminous than the non-starburst galaxies in our sample, we fit only these low-mass non-starburst galaxies to obtain a baseline of the $L_X-M_{200}$ relation, which can be characterized as a linear function:
\begin{equation}\label{equi:LXM200nonstarburst}
\frac{L_X}{10^{38}{\rm ergs~s^{-1}}}=(1.8\pm0.8)\frac{M_{200}}{{\rm 10^{12}M_\odot}}.
\end{equation}
In contrast, the average $L_X/M_{200}$ values of starburst and massive disc galaxies are $23.9_{-8.3}^{+14.6}$ and $34.2_{-10.5}^{+11.9}\times10^{26}\rm~ergs~s^{-1}~M_\odot^{-1}$, or about 13 and 19 times higher than that predicted by Eq.~\ref{equi:LXM200nonstarburst}. The GIMIC galaxies, however, exhibit a much steeper $L_X-M_{200}$ relation, although they are broadly consistent with the observed $L_X-v_{rot}$ relation (Fig.~\ref{fig:compareGIMIC}a). The $L_X-M_{200}$ relation of the GIMIC galaxies not only have a much steeper slope ($4.12\pm0.05$), but also have a much lower scatter ($rms=0.37\pm0.02\rm~dex$) than those of the observed galaxies ($rms=0.84\pm0.12\rm~dex$ for the low-mass non-starburst galaxies).

The $L_X-M_*$ correlation is poor for low-mass non-starburst galaxies, but can still be generally described with a linear function:
\begin{equation}\label{equi:LXMstarnonstarburst}
\frac{L_X}{10^{38}{\rm ergs~s^{-1}}}=(0.6\pm0.3)\frac{M_*}{{\rm 10^{10}M_\odot}}.
\end{equation}
Similarly, the average $L_X/M_*$ values of starburst and massive disc galaxies are much higher, i.e., $4.1_{-1.5}^{+2.4}$ and $11.5_{-3.4}^{+4.9}\times10^{28}\rm~ergs~s^{-1}~M_\odot^{-1}$, or about 7 and 19 times higher than that predicted by Eq.~\ref{equi:LXMstarnonstarburst}.

In contrast to the $L_X-M_{200}$ and $L_X-M_*$ relations, both the starburst and non-starburst galaxies in our \emph{Chandra} sample can be well described with an identical $L_X-{\rm SFR}$ relation (Fig.~\ref{fig:compareGIMIC}d), which can also be characterized with a linear function:
\begin{equation}\label{equi:LXSFR}
\frac{L_X}{10^{38}{\rm ergs~s^{-1}}}=(2.1\pm0.9)\frac{{\rm SFR}}{{\rm M_\odot~yr^{-1}}}.
\end{equation}
The coefficient here is only $\sim15\%$ of that for the $L_X-{\rm SFR}$ relation of the entire corona measured by our \emph{Chandra} observations (including the contribution from the inner region; Paper~II), but the $L_X-{\rm SFR}$ correlation is still good [Spearman's rank order coefficient (Paper~II) $r_s=0.62\pm0.13$]. The massive disc galaxies again show a large departure from the fitted relation, although they have SFRs comparable to the low-mass starburst galaxies. The observed galaxies with low SFRs (e.g., $\lesssim0.2\rm~M_\odot~yr^{-1}$) are typically dwarfs, which are excluded from the GIMIC sample. Besides these dwarf galaxies, the GIMIC galaxies still show significant differences from the observed sample in the $L_X-{\rm SFR}$ plane, i.e., the fitted $L_X-{\rm SFR}$ index of the GIMIC galaxies ($2.1\pm0.1$; for galaxies with ${\rm SFR}>0.5\rm~M_\odot~yr^{-1}$; the green solid line in Fig.~\ref{fig:compareGIMIC}d) is clearly larger than that of the observed sample (linear, Eq.~\ref{equi:LXSFR}). Furthermore, there are also some GIMIC galaxies with high coronal luminosities ($L_X\sim 10^{40}\rm~ergs~s^{-1}$) but just moderate SFRs ($\sim0.3\rm~M_\odot~yr^{-1}$), which are not seen in our observed galaxy sample. This is potentially an indication that massive galaxies are not being adequately quenched by feedback in the simulations.

\section{Discussion}\label{PaperIIIsec:Discussion}

\subsection{Effects of Missing AGN Feedback}\label{PaperIIIsubsec:AGNfeedback}

\begin{figure}[!h]
\begin{center}
\epsfig{figure=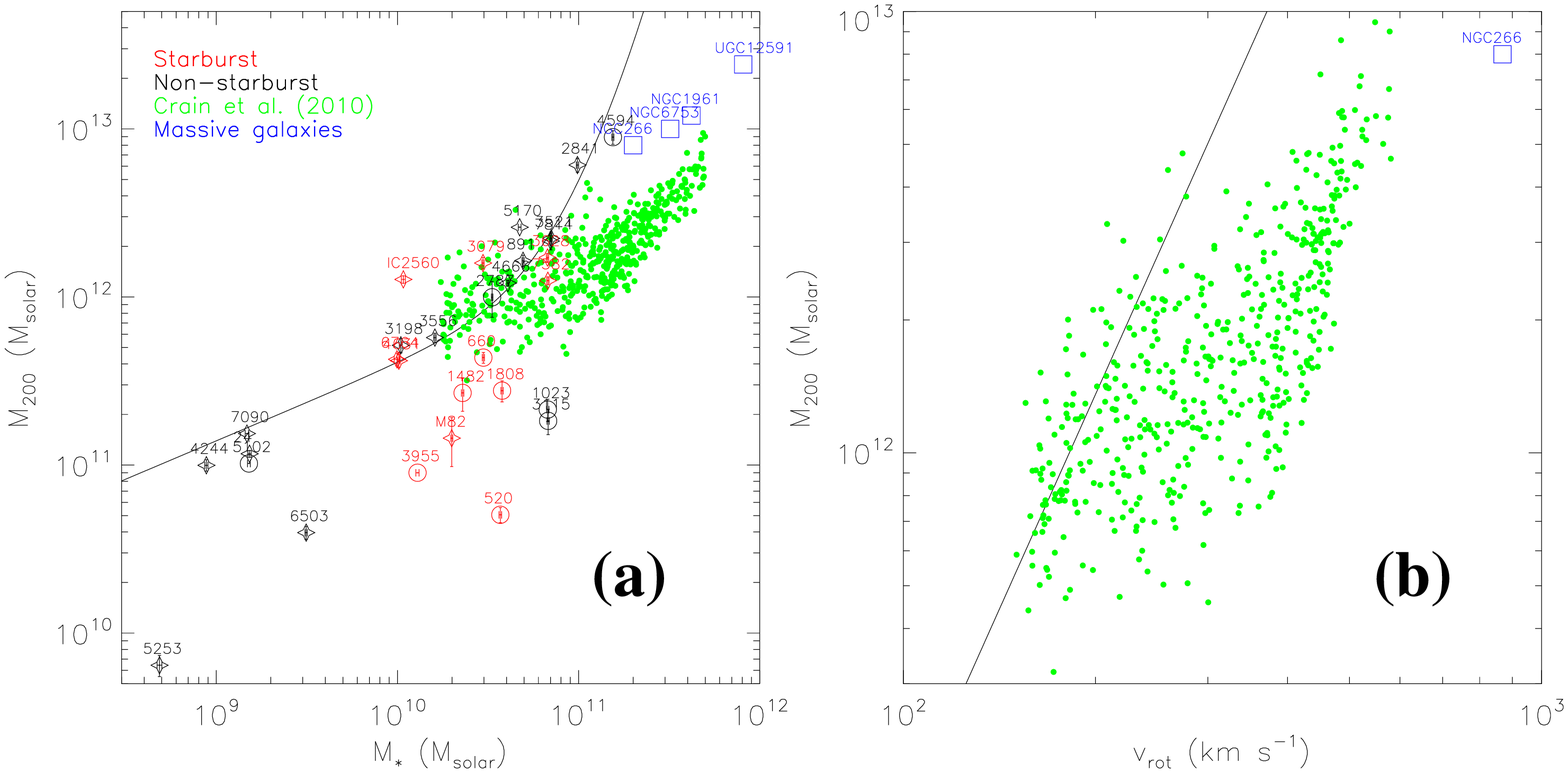,width=1.0\textwidth,angle=0, clip=}
\caption{(a) $M_{200}$ v.s. $M_*$. (b) $M_{200}$ v.s. $v_{rot}$. Symbols are the same as those in Fig.~\ref{fig:compareGIMIC}. The solid curve in panel (a) is the SMHMR from \citet{Leauthaud12a} [for the model $SIG\_MOD2$ (with the modeling of the stellar mass measurement errors) and the lowest redshift bin ($z=0.22-0.48$)]. The solid line in panel (b) marks the $M_{200}-v_{rot}$ relation of \citet{Navarro97}, which is used to compute $M_{200}$ of most of the observed galaxies, except for NGC~266.}\label{fig:M200Scaling}
\end{center}
\end{figure}

As introduced in \S\ref{PaperIIIsubsec:Simulations}, the GIMIC simulations do not include AGN feedback. The absence of AGN feedback in GIMIC may result in the over-prediction of the number density of \emph{massive} disc galaxies \citep{Crain09,Crain10a}. The detailed operation of AGN feedback on galaxy-wide scales remains ill-understood, but is expected to be important in galaxies with massive spheroids, where the energy output could heat and eject gas, reducing the emissivity in the central regions (e.g., \citealt{McCarthy10,McNamara12} and references therein). We have adopted a high-mass cut of $M_*\sim10^{11.7}\rm~M_\odot$ to the GIMIC galaxies to minimize the effect of missing AGN feedback (\S\ref{PaperIIIsubsec:Simulations}). Generally, this AGN effect is less important than Type~Ia SNe feedback for galaxies with $M_*\lesssim10^{11}\rm~M_\odot$ (e.g., \citealt{David06}), and should not qualitatively affect the comparison between observation and simulation in the mass range of our \emph{Chandra} sample. For galaxies with $M_*\gtrsim10^{11}\rm~M_\odot$, however, as shown in Fig.~\ref{fig:M200Scaling}a, the stellar masses at fixed $M_{200}$ are inconsistent with estimates derived from observational techniques \citep{Leauthaud12a}. This over-prediction of $M_*$ is most likely a result of the enhanced radiative cooling and SF in these massive galaxies due to the absence of AGN feedback in GIMIC. Correspondingly, this missing AGN feedback may also explain the high luminosity of coronal X-ray emission at a given $M_{200}$, for $M_{200}\gtrsim a~few\times10^{12}\rm~M_\odot$ (Fig.~\ref{fig:compareGIMIC}b).

To first order, the rotation velocity of galaxies is determined by the mass and concentration of their parent halo. Whilst the concentration of halos modeled in collisionless simulations is known to follow a tight relation (e.g., \citealt{Navarro97,Bullock01}), the co-evolution of baryons and dark matter in more realistic simulations induces systematic deviations and scatter in this relation (often described as a ``back reaction'', e.g., \citealt{Duffy10}). As shown in Fig.~\ref{fig:M200Scaling}b, low-mass galaxies ($M_{200}\lesssim (2-3)\times10^{12}\rm~M_\odot$) from GIMIC exhibit a larger scatter in $v_{rot}$ than more massive counterparts. The low-$v_{rot}$ boundary of low-mass galaxies is roughly consistent with \citet{Navarro97}'s relation (derived from collisionless simulations), and is used here to estimate $M_{200}$ of all the observed galaxies except for NGC~266. On the other hand, the high-$v_{rot}$ boundary of low-mass galaxies seems consistent with high-mass ones. In contrast, high-mass galaxies with $M_{200}\gtrsim (2-3)\times10^{12}\rm~M_\odot$ have clearly lower scatter in $v_{rot}$, and no galaxies have $v_{rot}$ as low as predicted by \citet{Navarro97}'s relation. In the absence of AGN feedback, the halo of a massive galaxy will tend to over-cool and over-contract, resulting in a greater rotation velocity. Therefore, the absence of AGN feedback in GIMIC, particularly for high-mass galaxies, likely results in an artificially high baryonic matter density at small radii, and thus a higher dark matter density and maximum rotation velocity.

\subsection{Role of SF in the Coronal X-ray Emission}\label{PaperIIIsubsec:CoolHotGasInteraction}

\begin{figure}[!h]
\begin{center}
\epsfig{figure=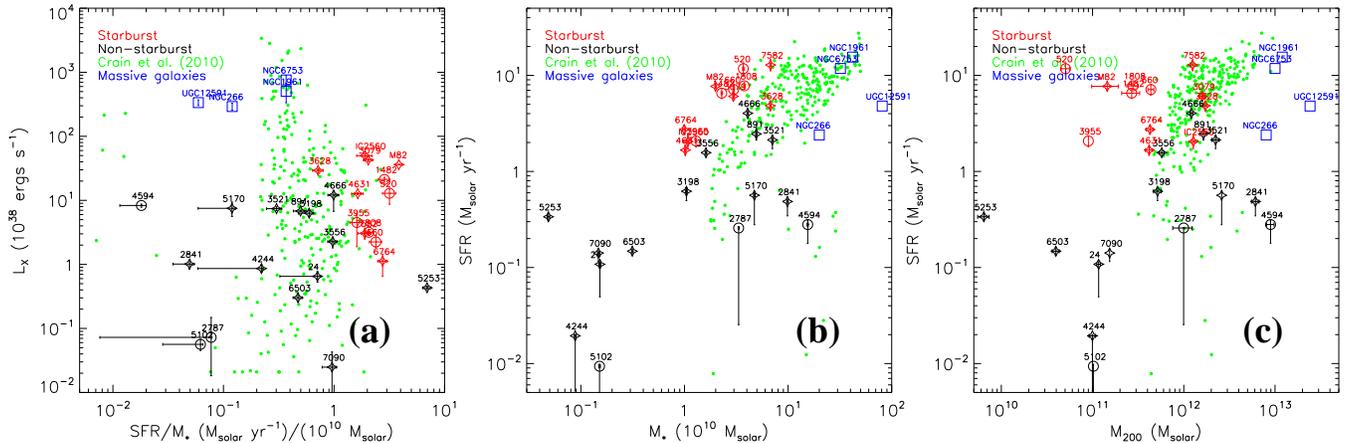,width=1.0\textwidth,angle=0, clip=}
\caption{SF properties of the sample galaxies. (a) $L_X$ v.s. the specific SFR per unit stellar mass (${\rm SFR}/M_*$). (b) SFR v.s. $M_*$. (c) SFR v.s. $M_{200}$. Symbols are the same as those in Fig.~\ref{fig:compareGIMIC}.}\label{fig:SFproperties}
\end{center}
\end{figure}

Our sample is selected based on the available \emph{Chandra} data. Archival observations are often proposed for active starburst galaxies which tend to have luminous X-ray halos. Among the 30 galaxies used in the present paper, $\sim40\%$ are classified as ``starburst'' ($\sim35\%$ if one includes the four massive disc galaxies; Table~\ref{table:GalaxyPara}), while in the GIMIC \textit{sample} (not necessarily the simulation taken as a whole), this fraction is only $\sim9\%$ (defining a simulated galaxy with ${\rm SFR}/M_*\gtrsim0.1\rm~Gyr^{-1}$ as a starburst, broadly consistent with the definition for the observed galaxies; Fig.~\ref{fig:SFproperties}a). This low fraction in the sample is likely a consequence of the exclusion from the sample of interacting galaxies. We thus separated starburst and non-starburst galaxies in the above comparisons (\S\ref{PaperIIIsubsec:compareGIMIC}).

The coronal luminosity discrepancy between the observed starburst and the GIMIC galaxies may be partly attributed to this difference in the sample selection. However, at small radii, the X-ray intensity always drops for GIMIC galaxies \citep{Crain10a,Crain13}, but increases for the observed ones (Paper~I). As presented in \S\ref{PaperIIIsubsec:compareGIMIC}, the coronal luminosity of our \emph{Chandra} sample galaxies in $0.01-0.1r_{200}$ is on average $\sim15\%$ of those adopted in Papers~I and II (\S\ref{PaperIIIsubsec:compareGIMIC}), while the corrected coronal luminosity of the GIMIC galaxies in $0.01-0.1r_{200}$ is typically $\sim(20-50)\%$ of the luminosity in the entire halo (\S\ref{PaperIIIsubsec:Simulations}). Considering that the X-ray luminosity adopted in Papers~I and II is also measured in the inner region, we could conclude that GIMIC under-predict the coronal X-ray emission of low-mass galaxies at least at small radii, although the chosen of photometry region of $r=0.01-0.1r_{200}$ has minimized this discrepancy. In fact, this under-prediction may also exist at $0.01-0.1r_{200}$, as indicated by the steeper slope of the $L_X-{\rm SFR}$ relation for the GIMIC galaxies (with respect to the observed galaxies in our \emph{Chandra} sample; Fig.~\ref{fig:compareGIMIC}d). We consider the SFR range of $0.5-5\rm~M_\odot~yr^{-1}$, because galaxies with larger SFRs are likely massive ones in GIMIC and starburst in our \emph{Chandra} sample, while those with a lower SFR in GIMIC have too large scatter in $L_X$ and many of them are indeed massive galaxies (see below; Fig.~\ref{fig:compareGIMIC}d). We find an average coronal luminosity of $\log (L_X/{\rm ergs~s^{-1}})=38.89_{-0.16}^{+0.15}$ for our observed galaxies, compared to $\log (L_X/{\rm ergs~s^{-1}})=38.02_{-0.10}^{+0.09}$ for the simulated ones.

The most intuitive reason for this under-prediction of X-ray luminosity around $L^\star$ non-starburst field galaxies may be from the adopted physical and/or numerical ingredients in GIMIC. Many physical assumptions and parameters, as well as the adopted numerical methods (e.g., \citealt{Scannapieco12}), may affect the simulated galaxy properties, e.g., the adopted feedback implementation, metal line cooling, stellar initial mass function, and the assumed cosmology (e.g., \citealt{Haas13a,Haas13b}). In particular, the choice of feedback parameters plays a key role in reproducing the observed SF properties of galaxies (e.g., \citealt{Schaye10,Scannapieco12,Vogelsberger13}), and possibly the related halo gas cooling and X-ray emission. As suggested in many cosmological simulations, there exists a fundamental problem that the specific SFRs of galaxies (particularly low-mass galaxies) at low-redshift are significantly under-predicted (e.g., \citealt{Weinmann12}). It has been suggested that this discrepancy arises because both semi-analytic and hydrodynamical simulations couple the growth of galaxies too strongly to the growth of their host dark matter haloes. Since haloes assemble early, simulated galaxies do so too, and thus form significantly earlier than is observed in nature (e.g., \citealt{Weinmann12}). Similarly, GIMIC also under-predicts the SFR of low-mass galaxies \citep{Crain09}. This under-prediction of current SF activity may have potential effects on the X-ray emission, e.g., via related stellar feedback and cool-hot gas interaction, but the net effect is not yet well understood. For example, there exists a small number of low-SFR high-$L_X$ galaxies in the GIMIC simulations (Fig.~\ref{fig:compareGIMIC}d). As indicated in Fig.~\ref{fig:SFproperties}b,c, these galaxies are most likely massive ones, which are apparently post-starbursts whose SF have already ceased but the radiative cooling of the coronal gas is still strong. Some observed galaxies (e.g., NGC~2787, NGC~2841, NGC~4594, and NGC~5170) may have similar SF properties, i.e., with low SFR but high $M_*$. The SF activities in these galaxies have probably been quenched, e.g., via morphological quenching or other mechanisms (e.g., \citealt{Martig09,Li09,Li11} and references therein). However, the diffuse X-ray luminosities of these galaxies are much lower than those of other massive galaxies with higher SFR (e.g., the massive disc galaxies from the literature; Fig.~\ref{fig:compareGIMIC}), apparently inconsistent with what is predicted by GIMIC.

The above comparisons clearly demonstrate that X-ray observations can place important constraints on the galaxy formation theory, especially its feedback implementation. Improved feedback prescriptions, especially in low-mass galaxies, may need to be considered in next generation simulations to correctly reproduce both the current SFR and the coronal X-ray luminosity. In particular, GIMIC adopted constant kinetic feedback parameters (initial wind velocity of $600\rm~km~s^{-1}$ and mass loading factor of $\dot{m}_{wind}/\dot{m}_*=4$), which were chosen to scale the global SFR density to correspond to observational estimates. Although the adopted efficiency (the wind energy accounts for $\sim80\%$ of the SN energy) and mass loading factor are roughly in agreement with observational estimates [e.g., for M82 by \citet{Strickland09}], it is likely that the macroscopic properties of outflows correlate with the detailed structure of the ISM (e.g., \citealt{Efstathiou00,Creasey13}) and, by extension, the properties of galaxies. The simplicity of the adopted physical and/or numerical ingredients in GIMIC is a possible reason that the SF and X-ray properties of the simulated galaxies are too strongly coupled with the dark matter halo mass of the galaxy (e.g., Fig.~\ref{fig:compareGIMIC}b). It will be interesting to compare the X-ray scaling relations of GIMIC galaxies with those of galaxies formed in next-generation hydrodynamical simulations adopting feedback schemes calibrated to reproduce the stellar mass function of local galaxies.

Another potential explanation for the under-prediction of soft X-ray emission is the adoption of a single-phase ISM in GIMIC. The gas is assigned an average temperature within the numerical resolution, which is typically insufficient to resolve a single massive SF region. In these SF regions, cold gas often dominates the total gas mass, while hot gas dominates the X-ray emission. Therefore, averaging the thermal state of the gases will result in a too low temperature to emit X-ray efficiently, thus significantly under-predict the X-ray emission. This effect is the most significant near a galactic disc, where the cool/hot gases strongly mix with each other, and the X-ray emission can be enhanced via enhanced radiative cooling of mass loaded cool gas and charge exchange (e.g., \citealt{Li11,Liu11,Liu12}). This effect may also help to explain the declining X-ray intensity profile toward small radii in GIMIC \citep{Crain10a,Crain13}, which is not typically observed in real galaxies.

As a general conclusion, GIMIC likely under-predict the role of SF in X-ray emission for low mass galaxies. There are two origins of this under-prediction: the under-prediction of the SFR for low-mass galaxies due to the too strong coupling of the growth of galaxies to the growth of their host dark matter halos, and the under-prediction of X-ray emission from SF due to the assumption of the single-phase ISM.

\subsection{Differences between Low and High Mass Galaxies}\label{PaperIIIsubsec:DifferenceLowHighMass}

Although the $L_X-M_{200}$ or $L_X-M_*$ slopes themselves are not well constrained due to the large scatter of the observational data, it is clear that the four massive disc galaxies (NGC~266, NGC~1961, NGC~6753, UGC~12591) are more X-ray luminous than the non-starburst galaxies below a typical transition mass of $M_*\sim2\times10^{11}\rm~M_\odot$ or $M_{200}\sim10^{13}\rm~M_\odot$ (Fig.~\ref{fig:compareGIMIC}b,c). As argued in previous studies (e.g., \citealt{Bogdan13a,Bogdan13b}), this high X-ray luminosity is most likely a signature of the accretion of intergalactic gas.

There are several possible mechanisms which may produce the high X-ray luminosity of these massive disc galaxies. \emph{Firstly}, the deeper gravitational potential, and possibly the stronger surrounding thermal and ram-pressure (e.g., \citealt{DallaVecchia08,Lu11}), may help the galaxy to retain more hot gas within its halo, resulting in an increasing baryon fraction (especially the hot gas fraction) with the halo mass (e.g., \citealt{Crain10a,Dai12}). \emph{Secondly}, the change of the $L_X/M_*$ ratio could also be a result of the change of the accretion mode. Numerical simulations have confirmed the change of the cold to hot accretion from below to above the (baryonic) mass of $\sim10^{10.3}\rm~M_\odot$ of a galaxy (e.g., \citealt{Keres05,Keres09,Crain10a,vandeVoort11}). Above this transition mass, a larger fraction of gas can be gravitationally heated to an X-ray emitting temperature in the galaxy vicinity. \emph{Finally}, the dynamical state of the coronal gas may also affect its overall X-ray emissivity by changing its radial distribution \citep{Ciotti91,OSullivan03}. A massive galaxy with a corona that tends to stay in a hydrostatic or even inflow state can have a steeper density profile (higher gas density in the center), enhancing the X-ray emission. Conversely, a galaxy with its corona in a subsonic outflow state tends to have a flatter density profile and may be less luminous in the galactic inner regions (e.g., \citealt{Tang09a}). In all cases, the dark matter halo mass appears to be a key parameter in determining the coronal luminosity and it is likely that the accretion of intergalactic gas plays an increasingly important role in producing the galactic coronae above the transition mass.

\section{Summary and Prospects}\label{PaperIIIsec:Summary}

In order to study the coronae around disc galaxies, we have constructed a \emph{Chandra} database of 53 nearby highly-inclined disc galaxies (Paper~I), and have conducted a correlation analysis of their coronal and other multi-wavelength properties (Paper~II). In this paper, we have compared our results to the predictions from a simple analytical model considering only the accretion of intergalactic gas (in the absence of feedback), the measurements of several massive disc galaxies from the literature, as well as results from recent cosmological hydrodynamical simulations invoking both accretion and feedback (the GIMIC simulations). Our main results and conclusions are set out below.

The observed X-ray emission in the vicinity of galactic discs (those explored in our \emph{Chandra} observations without further corrections in \S\ref{PaperIIIsubsec:OurSample}) is much less ($\sim2\%$) than predicted by a simple analytical model considering only the accretion of intergalactic gas. Furthermore, the radiative cooling rate of coronal gas in the same radial aperture is typically too low ($\lesssim10\%$ of the SFR) to be the primary source of gas to maintain the on-going SFR in the galaxies. The GIMIC simulations more accurately reproduce the observed X-ray scaling relations. They broadly reproduce the luminosity range of the coronal X-ray emission in $0.01-0.1r_{200}$ of $L^\star$ galaxies, including the scatter in $L_X$ at a given $v_{rot}$ or $M_*$. However, the $L_X-M_{200}$ relation of simulated galaxies differs from that inferred from observations, both in terms of the slope and scatter.

For the observed galaxies, low-mass starbursts appear to be more X-ray luminous than more quiescent counterparts at a given galaxy mass. There is a tight $L_X-{\rm SFR}$ correlation, even for corona at large radii of $r=0.01-0.1r_{200}$, similar to that of the X-ray emission at smaller radii. The overall trends of the observed and simulated galaxies on the $L_X-{\rm SFR}$ plot show little similarity, indicating that GIMIC potentially under-predicts the role of SF in producing the X-ray emission. There are two possible origins of this under-prediction: the under-prediction of the SFR for low-mass galaxies due to the too strong coupling of the growth of galaxies to the growth of their host dark matter halos, and the under-prediction of X-ray emission from SF due to the assumption of the single-phase ISM.

Coronal X-ray luminosity increases much faster with galaxy mass above a typical transition mass of $M_*\sim2\times10^{11}\rm~M_\odot$ or $M_{200}\sim10^{13}\rm~M_\odot$. Below this mass, the $L_X-M_{200}$ and $L_X-M_*$ relations of the observed galaxies can both be well characterized with a linear scaling relation. The higher $L_X/M_*$ or $L_X/M_{200}$ ratio of galaxies above the transition mass indicates that a massive disc galaxy tends to have a massive dark matter halo, to be dominated by hot-mode accretion, and/or to host a corona most likely in a hydrostatic state. The accretion of intergalactic gas likely plays an increasingly important role in producing the galactic coronae with increasing galaxy mass.

The above results demonstrate that X-ray observations can place important constraints on the galaxy formation theory, especially the astrophysics of accretion and feedback. Although the simulations have made a great progress to reproduce the observed X-ray properties of galaxies, some significant discrepancies still exist and are quite suggestive. The absence of AGN feedback in GIMIC tends to over-predict the number of massive galaxies, as well as a simultaneous over-prediction of $v_{rot}$, $M_*$, and $L_X$ at a given $M_{200}$ for massive galaxies. In addition, GIMIC may yield too strong a coupling of the galaxy and coronal properties ($v_{rot}$, $M_*$, SFR, $L_X$) to the host dark matter halo mass, which is one of the major origins of the discrepancies between the observations and simulations. Furthermore, the adoption of constant feedback parameters (mass loading and injection velocity) is likely too simplistic. It is also important to study the X-ray properties of low-mass galaxies, which are not massive enough to retain hot coronae, and the X-ray emission is mostly from stellar feedback and the related cool-hot gas interaction, which is poorly understood from current numerical simulations.

The present paper is based on the archival observations and X-ray measurements from the literature. Most of these observations are too shallow to probe the faint X-ray emission from the outskirts of the galactic halos. In addition, the sample is still not large nor uniform. In particular, very few galaxies are observed in X-rays around or above the transition mass. We have also focused on the \emph{X-ray luminosities} of the galactic coronae. Most of the archival observations are too shallow to yield valuable constraints on hot gas abundance, and hence on gas density and total baryon content of the galaxies (e.g., \citealt{Bregman07}). Therefore, deeper X-ray observations of more galaxies spanning the entire galaxy mass range are highly desirable in order to put tighter constraints on the galaxy formation models and further improve our understanding of the global properties of galactic coronae.

\acknowledgements

The authors thank $\rm\acute{A}$kos Bogd$\rm\acute{a}$n for helping us to compute the X-ray luminosity of NGC~266 in $0.01-0.1r_{200}$, Ying Zu and $\rm\acute{A}$kos Bogd$\rm\acute{a}$n for many helpful discussions, the anonymous referee for many useful suggestions that lead to improvements of the paper, as well as the referees of Papers~I and II whose constructive comments and suggestions led to the write-up of the present paper. Jiang-Tao Li acknowledges the financial support from CNES and the support from NSFC grant 11233001.

\end{document}